\documentclass[aps,prd,twocolumn,superscriptaddress]{revtex4-1}

\usepackage{aasdefs,graphicx}

\newcommand{\chead}[1]{\mbox{\bfseries #1}}

\newcommand{\anita}{\mbox{ANITA}}
\newcommand{\anitaone}{\mbox{ANITA-I}}
\newcommand{\anitatwo}{\mbox{ANITA-II}}
\newcommand{\anitathree}{\mbox{ANITA-III}}
\newcommand{\anitafour}{\mbox{ANITA-4}}

\newcommand{\eusospbtwo}{\mbox{EUSO-SPB2}}
\newcommand{\jemeuso}{\mbox{JEM-EUSO}}

\newcommand{\ice}{\mbox{IceCube}}
\newcommand{\icelong}{\mbox{IceCube Neutrino Observatory}}
\newcommand{\kmnet}{\mbox{KM3Net}}

\newcommand{\stau}{\mbox{$\tilde{\tau}_R$}}
\newcommand{\mstau}{{\tilde{\tau}_R}}

\newcommand{\nutausim}{\mbox{\tt NuTauSim}}

\newcommand{\aaei}{AAE\,061228}
\newcommand{\aaeiii}{AAE\,141220}
\newcommand{\icuptrk}{\mbox{IceCube-140611}}
\newcommand{\icuptwo}{\mbox{IceCube-140109}}
\newcommand{\icupthree}{\mbox{IceCube-121205}}

\newcommand{\enu}{\mbox{$\varepsilon_\nu$}}
\newcommand{\emu}{\mbox{$\varepsilon_\mu$}}
\newcommand{\emuobs}{\mbox{$\varepsilon_{\mu,{\rm obs}}$}}

\newcommand{\etau}{\mbox{$\varepsilon_\tau$}}
\newcommand{\etauobs}{\mbox{$\varepsilon_{\tau,{\rm obs}}$}}

\newcommand{\eproxy}{\mbox{$\varepsilon_{\rm proxy}$}}
\newcommand{\ecr}{\mbox{$\varepsilon_{\rm cr}$}}
\newcommand{\ecritt}{\mbox{$\varepsilon_{{\rm crit},\tau}$}}

\newcommand{\peveobs}{\mbox{$p(\varepsilon>\varepsilon_{\rm obs})$}}
\newcommand{\pzvzobs}{\mbox{$p(z>z_{\rm obs}\,|\,\varepsilon)$}}
\newcommand{\pjoint}{\mbox{$p_{\rm joint}$}}

\newcommand{\kmsq}{\mbox{km$^2$}}
\newcommand{\peryr}{\mbox{yr$^{-1}$}}
\newcommand{\kmsqsryr}{\mbox{km$^2$\,sr\,yr}}
\newcommand{\perkmsqsryr}{\mbox{km$^{-2}$\,sr$^{-1}$\,yr$^{-1}$}}

\usepackage{color}

\begin{document}

\title{The ANITA Anomalous Events as Signatures of a Beyond Standard
  Model Particle, and Supporting Observations from IceCube}

\author{Derek B. Fox}
\affiliation{Department of Astronomy \& Astrophysics, 525 Davey Lab, Penn State University, University Park, PA 16802, USA}
\email{dfox@psu.edu, steinn@psu.edu, nnp@psu.edu}
\affiliation{Center for Theoretical and Observational Cosmology, Institute for Gravitation and the Cosmos, 104 Davey Lab, Penn State University, University Park, PA 16802, USA}
\affiliation{Center for Particle and Gravitational Astrophysics, Institute for Gravitation and the Cosmos, 104 Davey Lab, Penn State University, University Park, PA 16802, USA}
\author{Steinn Sigurdsson}
\affiliation{Department of Astronomy \& Astrophysics, 525 Davey Lab, Penn State University, University Park, PA 16802, USA}
\affiliation{Center for Particle and Gravitational Astrophysics, Institute for Gravitation and the Cosmos, 104 Davey Lab, Penn State University, University Park, PA 16802, USA}
\author{Sarah Shandera}
\affiliation{Department of Physics, 104 Davey Lab, Penn State University, University Park, PA 16802, USA}
\email{shandera@psu.edu, murase@psu.edu, miguel@psu.edu, \\ coutu@phys.psu.edu}
\affiliation{Center for Fundamental Theory, Institute for Gravitation and the Cosmos, 104 Davey Lab, Penn State University, University Park, PA 16802, USA}
\affiliation{Center for Theoretical and Observational Cosmology, Institute for Gravitation and the Cosmos, 104 Davey Lab, Penn State University, University Park, PA 16802, USA}
\author{Peter M\'esz\'aros}
\affiliation{Department of Astronomy \& Astrophysics, 525 Davey Lab, Penn State University, University Park, PA 16802, USA}
\affiliation{Department of Physics, 104 Davey Lab, Penn State University, University Park, PA 16802, USA}
\affiliation{Center for Theoretical and Observational Cosmology, Institute for Gravitation and the Cosmos, 104 Davey Lab, Penn State University, University Park, PA 16802, USA}
\affiliation{Center for Particle and Gravitational Astrophysics, Institute for Gravitation and the Cosmos, 104 Davey Lab, Penn State University, University Park, PA 16802, USA}
\author{Kohta Murase}
\author{Miguel Mostaf\'a}
\author{Stephane Coutu}
\affiliation{Department of Physics, 104 Davey Lab, Penn State University, University Park, PA 16802, USA}
\affiliation{Department of Astronomy \& Astrophysics, 525 Davey Lab, Penn State University, University Park, PA 16802, USA}
\affiliation{Center for Particle and Gravitational Astrophysics, Institute for Gravitation and the Cosmos, 104 Davey Lab, Penn State University, University Park, PA 16802, USA}

\date{\today}


\begin{abstract}

  The \anita\ collaboration have reported observation of two anomalous events that appear to be $\ecr \approx 0.6$\,EeV cosmic ray showers emerging from the Earth with exit angles of $27\arcdeg$ and $35\arcdeg$, respectively. 
  While EeV-scale upgoing showers have been anticipated as a result of astrophysical tau neutrinos converting to tau leptons during Earth passage, the observed exit angles are much steeper than expected in Standard Model (SM) scenarios. 
  Indeed, under conservative extrapolations of the SM interactions, there is no particle that can propagate through the Earth with probability $p > 10^{-6}$ at these energies and exit angles. 
  We explore here whether ``beyond the Standard Model'' (BSM) particles are required to explain the ANITA events, if correctly interpreted, and conclude that they are.
  Seeking confirmation or refutation of the physical phenomenon of sub-EeV Earth-emergent cosmic rays in data from other facilities, we find support for the reality of the ANITA events, and three candidate analog events, among the Extremely High Energy Northern Track neutrinos of the IceCube Neutrino Observatory. 
  Properties of the implied BSM particle are anticipated, at least in part, by those predicted for the ``stau'' slepton (\stau) in some supersymmetric models of the fundamental interactions, wherein the stau manifests as the next-to-lowest mass supersymmetric partner particle. 
  
\end{abstract}

\maketitle


\section{Introduction\label{sec:intro}}

The Antarctic Impulsive Transient Antenna (\anita) Collaboration recently revisited and expanded upon their earlier report \cite{gnr+16} of an anomalous upgoing air shower from the \anitaone\ flight, reporting a second upgoing air shower from \anitathree\ \cite{gra+18} and exploring possible interpretations for these events (see also \cite{mapz17,cs18,abl+18,huang18,cab18}).

Transient ($\delta t \approx 6$\,ns) radio-frequency ($30\,{\rm  MHz}\simlt \nu\simlt 1\,{\rm GHz}$) pulses due to ultra-high energy cosmic rays (UHECRs; $\ecr\simgt 10^{17}$\,eV) were regularly observed during radio-quiet periods of the \anitaone\ and \anitathree\ long-duration balloon flights over the Antarctic continent \cite{hng+10,sbr+16,gnr+16,gra+18} (the trigger algorithm used for \anitatwo\ was not sensitive to these events \cite{gnr+16}). 
Over 90\% of these showers exhibit inverted polarity relative to the local geomagnetic field; such (non-anomalous) showers are explained by development of a UHECR shower in the upper atmosphere elsewhere over Antarctica, producing a radio pulse of angular extent $\delta\theta\approx 3\arcdeg$ \cite{hng+10} which reflects off the Antarctic ice cap before detection by \anita. 
The expected radio properties of the showers, along with development and demonstration of an energy calibration for the events, are presented in \cite{sbr+16}.

By contrast, the two \anita\ anomalous events (hereafter AAEs) exhibit non-inverted polarity, consistent with geomagnetic effects for a shower observed directly by \anita, without reflection. 
Because of the experiment's $>$30\,km altitude during flight, direct observation of atmospheric CR showers is possible within $\delta z \le 6\arcdeg$ of the horizontal, close to but above the physical horizon. 
A few such events are seen, and are not considered anomalous. 
The two anomalous showers, by contrast, are incident from well below the horizon, having zenith angles of $z'=117\fdg 4 \pm 0\fdg 3$ (\anitaone) and $z'=125\fdg 0 \pm 0\fdg 3$ (\anitathree), respectively (Table~\ref{tab:evts}). 
They are interpreted as due to upgoing (Earth-emergent) cosmic ray showers with energies of $\ecr \approx 0.6$\,EeV each.


\begin{table*}
\caption{Properties of the ANITA Anomalous Events\label{tab:evts}}
\begin{ruledtabular}
  \begin{tabular}{lcc}
    \chead{Property} & \chead{AAE\,061228}
                     & \chead{AAE\,141220} \\ \hline
    Flight \& Event  & \anitaone\ \#3985267
                     & \anitathree\ \#15717147 \\
    Date \& Time (UTC) & 2006-12-28 00:33:20
                       & 2014-12-20 08:33:22.5 \\
    Equatorial coordinates\ (J2000) & R.A. 282\fdg 14064, Dec.\ +20\fdg 33043
                     & R.A. 50\fdg 78203, Dec.\ +38\fdg 65498 \\
    Energy \ecr      & $0.6\pm 0.4$\,EeV
                     & $0.56^{+0.30}_{-0.20}$\,EeV \\
    Zenith angle $z'$/$z$ & $117\fdg 4$ / $116\fdg 8 \pm 0\fdg 3$
                          & $125\fdg 0$ / $124\fdg 5 \pm 0\fdg 3$ \\
    Earth chord length $\ell$ & $5740\pm 60$\,km
                              & $7210\pm 55$\,km \\
    Mean interaction length for $\enu=1$\,EeV
                        & 290\,km    
                        & 265\,km \\ 
    $p_{\rm SM}(\etau > 0.1\,{\rm EeV})$ for $\enu=1$\,EeV
                     & $4.4 \times 10^{-7}$
                     & $3.2 \times 10^{-8}$ \\
    $p_{\rm SM}(z > z_{\rm obs})$ for $\enu=1$\,EeV, $\etau > 0.1$\,EeV
                     & $6.7 \times 10^{-5}$
                     & $3.8 \times 10^{-6}$ \\
    $n_\tau(\mbox{1--10\,PeV})$
                     : $n_\tau(\mbox{10--100\,PeV})$
                     : $n_\tau(>0.1\,\mbox{EeV})$
                     & 34 : 35 : 1
                     & 270 : 120 : 1 \\
  \end{tabular}
\end{ruledtabular}
\end{table*}


While this interpretation is straightforward, it raises sharp challenges within an SM framework. 
We explore these challenges below, but in brief: No SM particle is expected to survive passage through the Earth, along associated chord lengths $\ell > 5700$\,km, at such energies. 
In particular, while a UHE tau neutrino ($\nu_\tau$) can ``regenerate'' from $\tau$ decays \cite{bdp+08} and convert to a $\tau$ shortly before or after emergence, producing a UHE upgoing air shower, incident EeV $\nu_\tau$ on such trajectories yield $\etau>0.1$\,EeV emerging tau leptons with probabilities $p_{\rm SM} < 10^{-6}$ (Table~\ref{tab:evts}). 
Attempting to accommodate the AAEs in a strictly SM scenario, \cite{gra+18} suggest that gluon saturation might lead the SM neutrino cross section to plateau (rather than continue to increase) above $\enu\sim 10^{18}$\,eV. 
As they concede, even in this case, existing constraints on diffuse UHE neutrino fluxes \cite{auger15_nu,ic18_diffnu} mean that bright and impulsive UHE neutrino-emitting transients would likely be required to explain the AAEs, raising other difficulties.

Our paper proceeds as follows. 
In Sec.~\ref{sec:sm} we review the evidence against SM explanations for the AAEs, introducing new arguments more definitive than those presented to date, and addressing the implications of our findings. 
We argue that existence of a BSM particle with appropriate properties would resolve all outstanding questions regarding the otherwise extremely unlikely properties of these events. 
Given the strong exclusion of SM scenarios, and the promising case for BSM explanations, we proceed in Sec.~\ref{sec:other} to seek out confirming or refuting observations from other facilities. 
We find that the \icelong\ holds the greatest promise in this regard.
Reviewing the highest-energy neutrino events from \ice, we find that these already provide independent support for the reality of the AAE phenomenon, including three candidate analog events. 
In Sec.~\ref{sec:theory} we review theoretical precedents for the anomalous \anita\ and \ice\ events in the literature, and implications of these theories. 
In Sec.~\ref{sec:conclude} we conclude that, taken together, the \anita\ and \ice\ anomalous events provide dramatic and highly credible evidence of the first new bona fide BSM phenomenon since the discoveries of neutrino oscillations, dark matter, and dark energy. 


\section{Upgoing Showers in the Standard Model\label{sec:sm}}

The AAEs are forbidden under strictly SM scenarios on at least two grounds.

\subsection{Diffuse neutrino flux limits}

The AAE trajectories are highly improbable. 
Any SM-based estimate for the optical depths along these trajectories, for incident neutrino energies greater than the estimated shower energies, leads to implied neutrino fluxes well in excess of published bounds from the Pierre Auger Cosmic Ray Observatory \cite{auger15_nu} and the \icelong\ \cite{ic18_diffnu}. 
Because of the potential for $\nu_\tau$ regeneration effects to complicate neutrino survival calculations, and the requirement that events generate a high-energy tau near Earth's surface to be observed, we have explored this question via simulations; specifically, the \nutausim\ $\nu_\tau$ propagation and emergent $\tau$ shower software of \cite{acp+18}.

We begin by injecting a monoenergetic flux of 100~million $\nu_\tau$ at each half-decade of energy between 0.1\,EeV and 1000\,EeV (nine energies in all), along the trajectories of each AAE (1.8~billion total injections). 
\nutausim\ propagates each neutrino according to SM physics, allowing for a choice of $\tau$ energy loss models and neutrino cross sections beyond $\enu > 0.1$\,EeV, and using the Preliminary Reference Earth Model (in a purely spherical approximation) from \cite{earth_ref}. 
During propagation the $\nu_\tau$ will typically participate in multiple charged current (CC) and neutral current (NC) interactions with nucleons; the $\tau$ particles generated by CC reactions propagate, lose energy, and decay, regenerating (at lower energy) the $\nu_\tau$. 
Propagation stops when the energy of the primary $\nu_\tau$ or $\tau$ particle drops below 0.1\,PeV, as the authors were concerned with UHECR shower observations with current and near-future facilities \cite{acp+18}, for which this energy falls well below threshold. 
Energies and interaction histories of $\tau$ particles that successfully emerge from the surface of the Earth with $\etau > 0.1$\,PeV are recorded for analysis, while $\nu_\tau$ that emerge from the Earth are assumed to escape undetected.

\begin{figure*}
  \begin{centering}
    \includegraphics[width=0.5\textwidth]{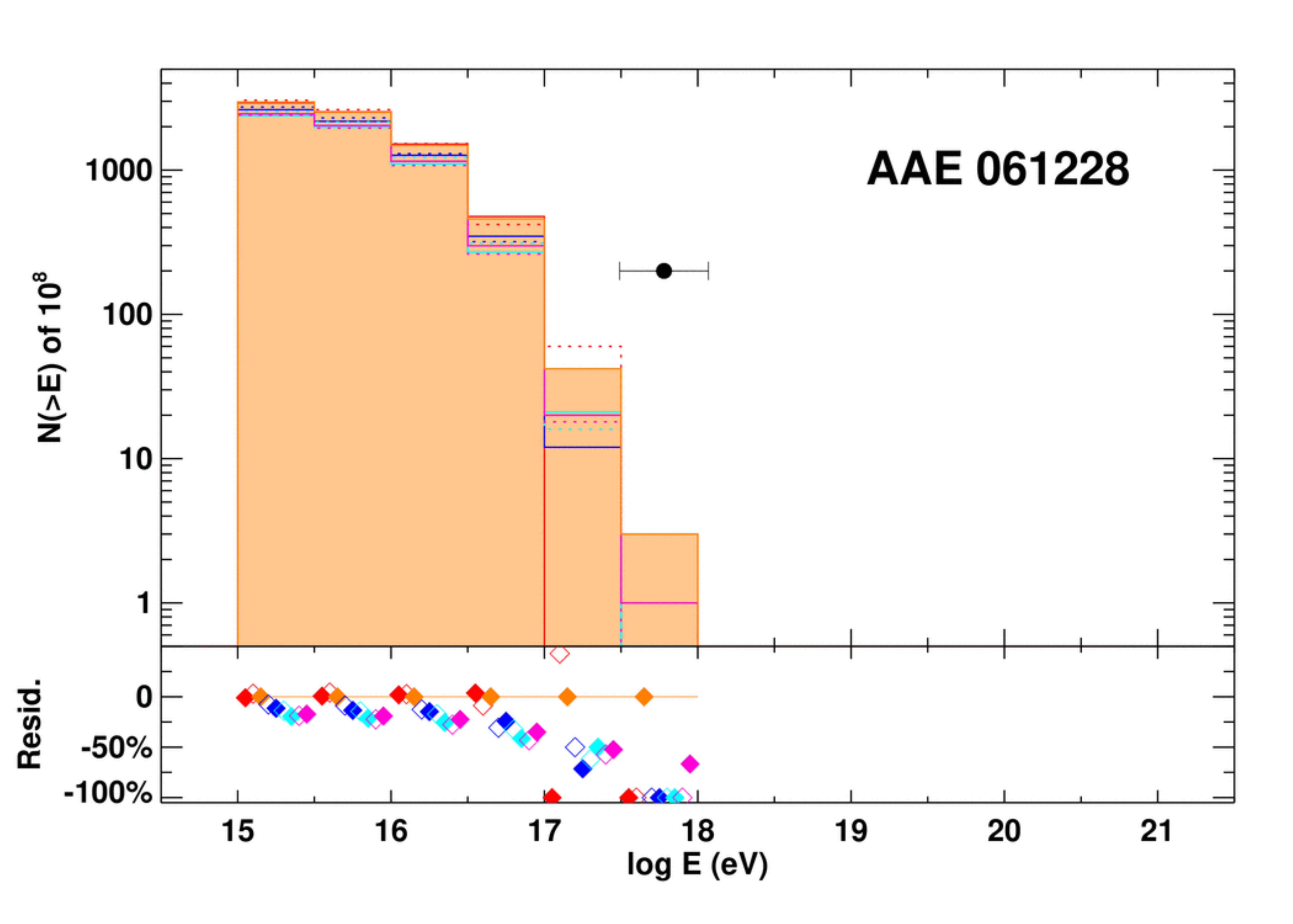}%
    \includegraphics[width=0.5\textwidth]{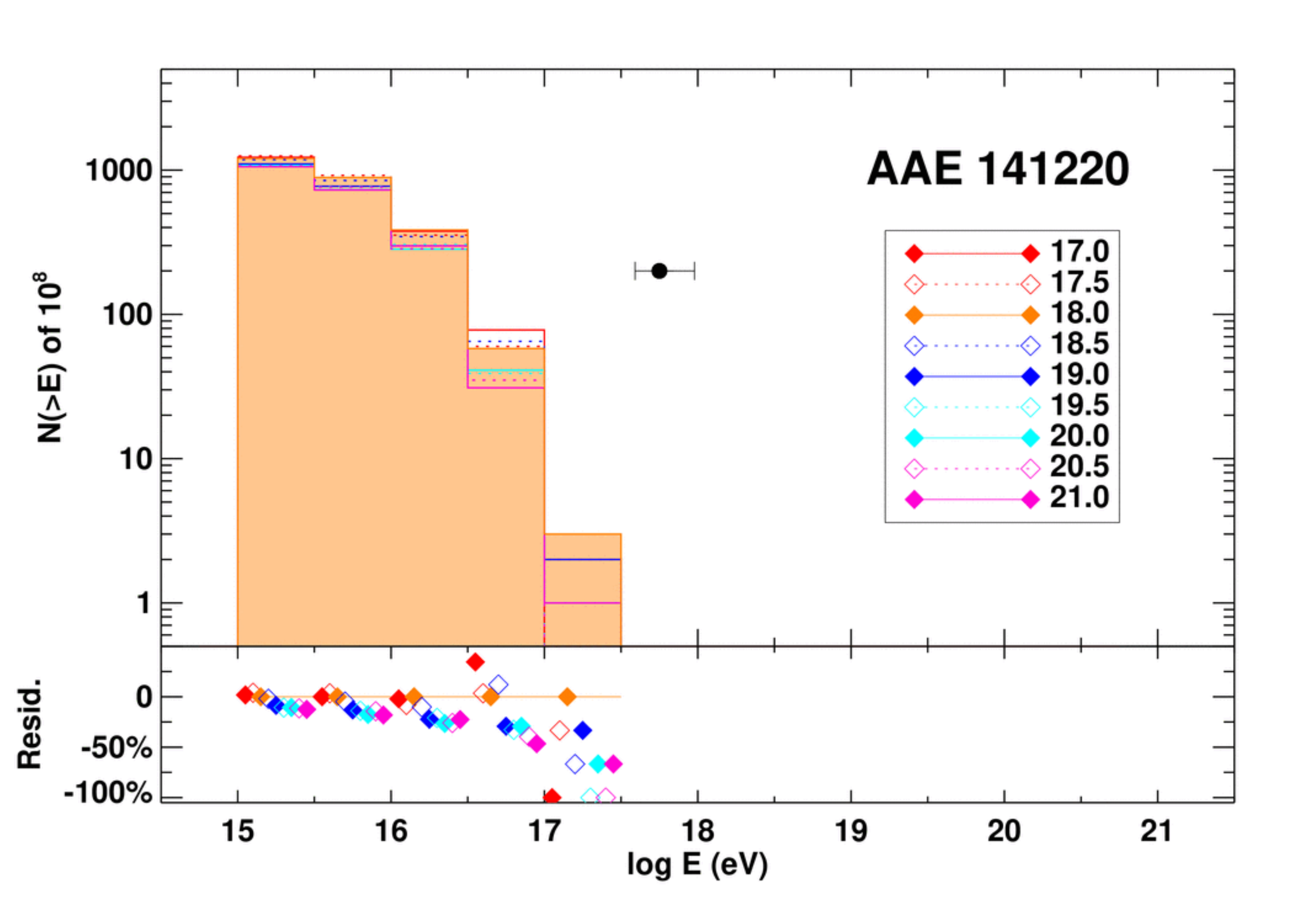}
  \end{centering}
\caption{Cumulative histograms (top panels) of observed tau energies for injected tau neutrinos over a range of energies from 0.1\,EeV to 1000\,EeV (indicated by legend in right panel, labeled by the logarithm of the energy in eV) for the two \anita\ anomalous events AAE\,061228 (left) and AAE\,141220 (right). 
Observed energies of the events are indicated by the black dots (with error bars). 
Histograms show the number of Earth-emergent tau following propagation of 100~million tau neutrinos through the Earth, at or greater than the lower energy boundary for each bin, along the trajectory of each event; the histogram for input neutrino energy $\enu=1$\,EeV is shaded in light orange and serves as a reference for the residuals plot below. 
Bottom panel: Residuals of the cumulative distribution for each energy, compared to the distribution for $\enu=1$\,EeV. 
For both trajectories, $\enu=1$\,EeV neutrinos provide a near-maximal number of emergent $\etau>0.1$\,EeV tau particles. \label{fig:etau}}
\end{figure*}

The resulting distribution of energies for Earth-emergent $\tau$ (Fig.~\ref{fig:etau}) shows a sharp cutoff between 0.1\,EeV and 1\,EeV that holds regardless of input neutrino energy: the maximum-energy emergent $\tau$ from the two sets of simulations has $\etau = 0.71$\,EeV (for \aaei) and 0.49\,EeV (for \aaeiii), respectively. 
The approximate maximum energy follows from the dominance of photonuclear energy losses for $\tau$ propagating through dense media, which are proportional to the particle energy \cite{acp+18}. 
As a result, $\tau$ with energies exceeding $\ecritt \approx 0.34$\,EeV in ice (0.12\,EeV in rock) lose energy rapidly -- typically before decaying -- until reaching \ecritt.


Since all successful trajectories involve production of a high-energy $\tau$ in ice or rock, it is not possible to observe $\etau > \ecritt$ emerging tau along these deep trajectories. 
This is a potentially important point in understanding the observed energies of the AAEs, which we will return to later. 
For the present argument, it will suffice to note the following two points.

First, the simulations demonstrate that neutrinos of energy $\enu \sim 1$~EeV are the most likely to yield high-energy emerging tau particles from deep trajectories. 
The reason is that $\enu > 1$\,EeV neutrinos have higher cross sections (even in the ``lower'' cross section model of \cite{acp+18}), leading to earlier interaction and tau production. 
The greater energy of this first tau is quickly lost, however, as photonuclear losses bring it down to \ecritt, just as for the first tau from a lower-energy EeV neutrino. 
Subsequent to first tau production, then, effectively all trajectories with $\enu > 1$\,EeV yield $\etau \approx \ecritt$ propagating taus.

Second, and for this reason, details of neutrino cross sections at $\enu > 1$\,EeV -- still somewhat uncertain under the SM
\cite{gra+18,acp+18} -- are very nearly irrelevant. 
As long as the total neutrino-nucleon cross section does not decrease substantially below its value at 1~EeV, any such differences will change only the mean distance to first tau conversion, a minor effect in the context of these extended path lengths. 

With minimal loss of generality, then, henceforward we explore the success rate for injected 1\,EeV neutrinos along the AAE trajectories, and other trajectories, using the standard SM neutrino cross sections from \cite{ctw11}. 
Furthermore, in this section, we treat emergent tau with $\etau \ge 0.1$\,EeV as detectable by \anita. 
This choice is conservative in that it should favor SM explanations, which (as we have seen) are only barely capable of generating emergent tau with the quoted ``best fit'' energies of the AAEs along deep trajectories, and generate a significantly greater number of emergent taus down to a 0.1~EeV threshold.

We have not explored the variation in results due to different inelasticity distributions for  neutrino interactions at these energies. 
Successful emergent $\etau \ge 0.1$\,EeV tau following the AAE trajectories are predominantly first- or second-generation tau particles. 
As such, increasing (or decreasing) the mean inelasticity in CC interaction by 25\% would increase (decrease) the total energy lost prior to tau emergence by at most 50\%. 
We note that \ice\ recently validated SM models for the inelasticity of neutrino CC interactions at sub-PeV energies \cite{ic18_inelastic}. 

We calculate the success rate for 1\,EeV tau neutrinos incident along each of the two AAE trajectories by simulating 200~million (\aaei) and 1~billion (\aaeiii) such injections, respectively.
Results are presented in Table~\ref{tab:evts}: Probabilities for success are $(4.4 \pm 0.5) \times 10^{-7}$ and $(3.2 \pm 0.6) \times 10^{-8}$ for \aaei\ and \aaeiii, respectively. 

We estimate the exposures of \anitaone\ and \anitathree\ below; we find a total exposure of 2.7\,\kmsqsryr\ based on the number and energies of the reflected UHECR events detected during each flight. 
Detection of successful emergent tau from two EeV tau neutrinos at these zenith angles, then, implies a rate of incident tau neutrinos with $\enu > 1$\,EeV of roughly 12~million\,\perkmsqsryr; this is over a million times in excess of
the Pierre Auger Observatory \cite{auger15_nu} and \ice\ \cite{ic18_diffnu} bounds, both of which give $\enu\,\phi(\enu) \simlt 6$\,\perkmsqsryr\ at $\enu=1$\,EeV.

In order to evade this constraint, \cite{gra+18} propose that the AAEs might be generated by extremely high-luminosity EeV neutrino
transients. 
This hypothesis can indeed evade the \ice\ and Auger diffuse bounds; however, any cosmic population of such transients (assumed extragalactic) must be isotropic, ongoing, and include sources exhibiting a broad range of fluxes here at Earth, due both to source distance effects and any associated luminosity function. 
No such cosmic population of high-luminosity, high-frequency
($\gg$2~month$^{-1}$, all-sky) neutrino transients can be compatible with the limits on neutrino point sources \cite{ic17_icrc17}, rates of $\enu\simgt 200$\,TeV \cite{ic17_icrc17} and $\enu\simgt 1$\,PeV \cite{ic18_diffnu} neutrinos, and rates of neutrino multiplet events \cite{ic17_multiplet} set by \ice.

An SM explanation for the two AAEs is thus ruled out by existing diffuse neutrino background limits, given the extreme improbability of success for individual neutrinos along these trajectories. 
To quantify this statement, we note that the Poisson probability of detecting two or more events against an expectation of $\lambda = 2.4\times 10^{-6}$ events (upper bound on diffuse flux, divided by three for tau only, times success rate for \aaei, times total \anita\ exposure) is $p_{\rm diffuse} = 2.8\times 10^{-12}$, which excludes SM scenarios at 7.0$\sigma$ confidence.


\subsection{ANITA zenith angles}

Independent of the sheer improbability of the AAEs, the two observed zenith angles are highly improbable under the SM.
The surprising steepness of the AAE trajectories has been noted in previous treatments \cite{gnr+16,cs18,abl+18,huang18,cab18}.

\begin{figure}
  \begin{centering}
    \includegraphics[width=0.5\textwidth]{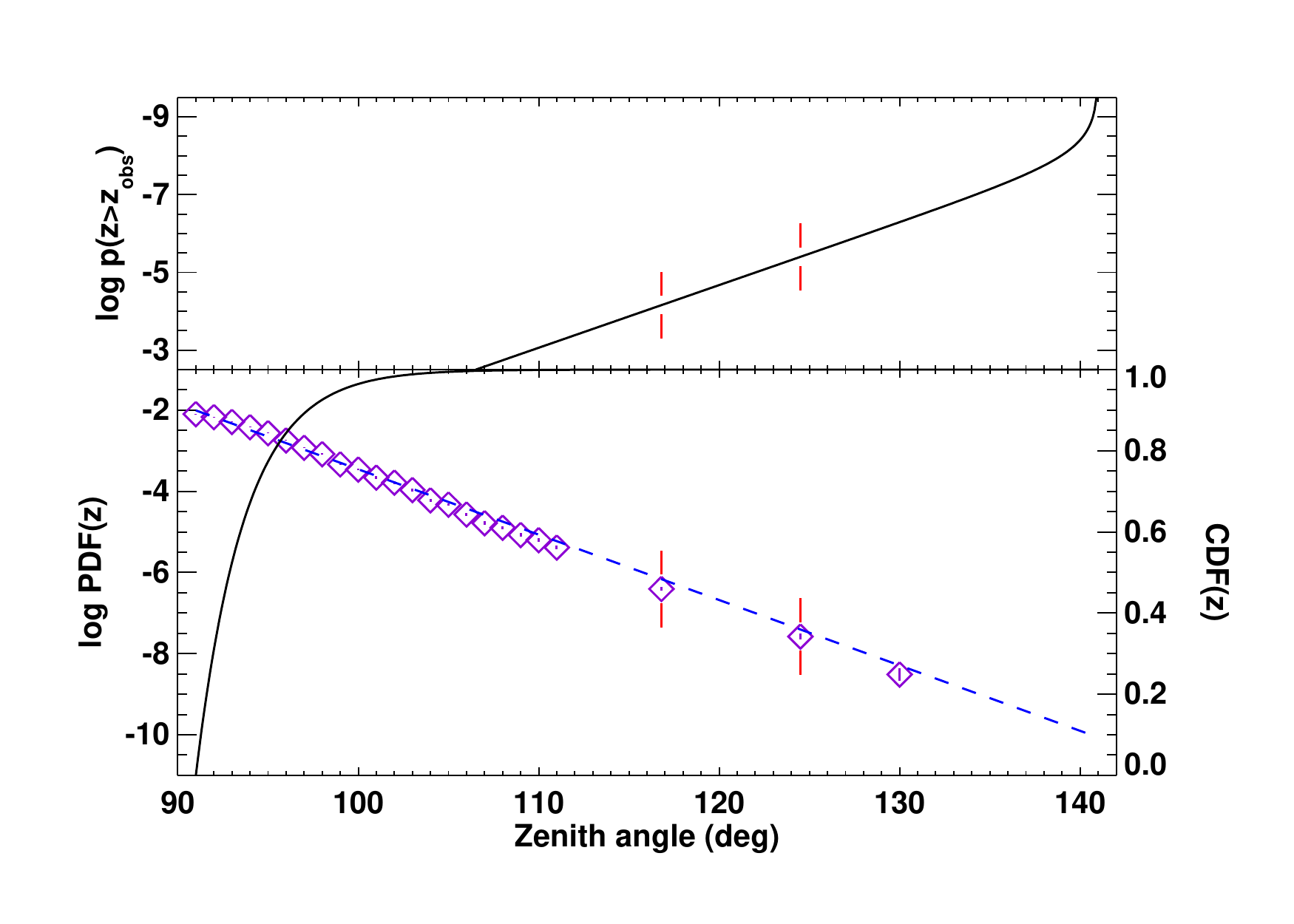}
  \end{centering}
\caption{Expected zenith angle distribution for AAEs under the SM given an isotropic flux of incident $\enu \simgt 1$\,EeV neutrinos. 
  We calculate the expected distribution by multiplying the success rate at any given zenith angle ($z$) by $\sin z$ to account for solid angle effects. 
  We fit an exponential model to the data and extrapolate to a maximum zenith angle $z_{\max} = 141\arcdeg$.
  Zenith angles (red hash marks) and simulation results for \aaei\ and \aaeiii\ are indicated. 
  Bottom panel: Exponential distribution in zenith angle as a PDF (blue dashed line), showing simulations data (purple diamonds), and the associated cumulative distribution (black line, with scale provided on the right). 
  Top panel: Log of the CDF residual, showing the $p$-values for the two AAEs. \label{fig:zenith}}
\end{figure}

To quantify this effect, we run simulations to calculate the success rate as a function of zenith angle for incident $\enu=1$\,EeV tau neutrinos, with success defined (as before) as events yielding an emergent $\etau \ge 0.1$\,EeV tau. 
We find that the success rate, as weighted by solid angle, declines exponentially with zenith angle (Fig.~\ref{fig:zenith}), with $e$-folding angular distance $\delta z = 2\fdg 7$. 
We approximate this distribution as a pure exponential and construct the probability distribution function (PDF) and its cumulative function (the CDF) for AAEs in zenith angle, under the SM, over $91\arcdeg \le z \le 141\arcdeg$.

As an aside, we note that the zenith angle $z'$ observed by \anita\ is not identical to the zenith angle $z$, relative to Earth's surface, reflecting the particle's trajectory through the Earth (e.g., for purposes of \nutausim\ simulation), due to the combined effects of the balloon altitude ($h\approx 35$\,km) and Earth curvature.  
We have calculated and use corrected $z$ values for the AAEs as shown in Table~\ref{tab:evts}.


We choose a minimum zenith angle of $91\arcdeg$ because a nonzero path length through dense media is required to realize the first neutrino interaction; we choose a maximum zenith angle because the sensitivity of the \anita\ experiment does not extend to the nadir \cite{anita09_instrument,hng+10}; it must extend to at least $z = 124\fdg 5$ given observation of the \aaeiii\ event. 
We make a conservative choice of $z_{\max} = 141\arcdeg$ which is $45\arcdeg$ beyond the \anita\ horizon ($z'=96\arcdeg$ \cite{gra+18}). 
As this bound lies more than six $e$-foldings beyond the largest observed angle, there will be negligible integrated probability density at even greater angles. 
Over this range of zenith angle, we assume the \anita\ detectors deliver uniform sensitivity.

We use the normalized ${\rm CDF}(z)$ to calculate a $p$-value for each AAE, defined as the chance for \anita\ to observe a successful emergent tau event at the specific zenith angle or beyond. 
We find $\pzvzobs=6.7\times 10^{-5}$ for \aaei\ and $\pzvzobs=3.8\times 10^{-6}$ for \aaeiii\ (Fig.~\ref{fig:zenith}). 
Combining the two $p$-values by Fisher's method, we find a joint $p_z = 5.9\times 10^{-9}$, excluding the SM-derived distribution at 5.8$\sigma$ confidence.


\subsection{Other considerations}

We can exclude SM scenarios for the AAEs on multiple further grounds by referring to limits from other facilities on UHE neutrinos and UHE neutrino-emitting transients. 
We address the relative exposures of other facilities in Sec.~\ref{sec:other}; since we conclude that \ice\ provides the strongest constraints, for present purposes we restrict our attention to \ice, referring forward to these results.

\ice\ offers a total exposure to UHE tau-initiated hadronic showers that we estimate at 54.0\,\kmsqsryr, assuming an isotropic distribution over the Northern hemisphere where \ice\ has nearly zero background. 
This exposure can be compared to an estimated upper bound on the total exposure for \anita\ (also estimated below) of 2.7\,\kmsqsryr. 

The expected number of events for \ice\ is thus roughly an order of magnitude more than for \anita. 
As with \anita, SM explanations for the AAEs will founder on the absence of UHE neutrino detections, and (for explanations invoking transient source populations) neutrino multiplet events, at reduced zenith angles in \ice, where UHE neutrino passage, and generation of UHE emergent taus, is exponentially more likely. 
In this context, note that the recent nine-year search for $\enu\simgt 1$\,PeV neutrinos in \ice\ identified just two events \cite{ic18_diffnu}.

Moreover, as also illustrated in \cite{ic18_diffnu}, \ice\ retains high sensitivity to tau-induced showers down to $\enu\sim 1$\,PeV before encountering a noticeable background from UHECR shower-generated neutrinos. 
Since $\enu\sim 1$\,EeV tau neutrinos incident along deep trajectories produce many more emergent tau at lower energies than above the $\etau \approx 0.1$\,EeV \anita\ threshold (Fig.~\ref{fig:etau}; Table~\ref{tab:evts}), even monoenergetic astrophysical sources of $\enu = 1$\,EeV tau neutrinos can be excluded in this fashion, given the absence of lower-energy events in the \ice\ dataset.

Finally, due to neutrino oscillations en route to Earth, astrophysical sources are generally expected to produce equal fluxes of the three neutrino flavors. 
Since \ice\ exhibits substantially greater sensitivity to electron antineutrinos near the Glashow resonance at $\enu=6.3$\,PeV \cite{glashow60}, and also provides greater sensitivity to $\enu > 1$\,PeV muon neutrinos, given the many-km range of such muons in ice \cite{ic16_n_muon}, the absence of large numbers of Glashow resonance and $\emu > 200$\,TeV muon events in the \ice\ dataset
\cite{ic16_n_muon,ic18_diffnu}, if quantitatively evaluated, would likely also rule out SM scenarios with high confidence.


\subsection{SM exclusion and implications\label{sub:smexclude}}

We have ruled out SM scenarios for the AAEs, by two distinct lines of argument, at confidence levels of 7$\sigma$ and 5.8$\sigma$, respectively. 
This evidence level is sufficient to support a claim of a new BSM phenomenon, if the interpretation of the AAEs as UHE upgoing air showers at large zenith angle can be confirmed from other facilities. 

We wish to be clear about the nature of this exclusion. 
After all, there are certainly $\enu \sim 1$\,EeV tau neutrinos incident on Earth at some flux level, due to astrophysical and cosmogenic production processes.
These UHE tau neutrinos are highly penetrating ($\sigma_{\rm tot} \sim 15$\,nb at 1\,EeV), and participate in the ``tau regeneration'' process we have simulated, which degrades the energy of the primary particle only modestly during each generation, and readily converts $\enu \simgt 1$\,EeV tau neutrinos to $\etau \simlt \ecritt \sim 0.3$\,EeV tau leptons during Earth passage over modest path lengths ($\ell \simlt 500$\,km).
As we have found, however, the chord lengths and column densities for \aaei\ and \aaeiii\ are simply too great to accommodate successful tau emergence from these trajectories within the SM (Table~\ref{tab:evts}). 

As such, the key missing element needed to construct a self-consistent physical picture for the production, propagation, and detection of the AAEs is an intermediary BSM particle, produced in UHECR interactions, with a substantially lower cross section to nuclear scattering and (if charged) minimal electromagnetic energy losses during propagation.
Its lifetime should be sufficient to propagate across the Earth, $\tau_{\rm BSM} \simgt 10\,(m_{\rm BSM}/500\,{\rm GeV})$\,ns, following which it should convert, via decay or interaction, into a tau lepton or tau neutrino. 
Once passage across the first $\approx$5000\,km of the trajectory is complete, and the BSM particle has produced a UHE tau or tau neutrino, the physics of tau propagation and regeneration ensure a reasonable probability for successful emergence of an $\etau \simlt 0.3$\,EeV tau. 
Hadronic or electromagnetic decay of this tau in the lower atmosphere will then yield an upgoing UHECR shower and associated radio pulse, as observed by \anita. 

In order that its existence not be excluded by existing searches \cite{atlas18_dilepton,cms18_tau}, the BSM particle should probably have a mass $m_{\rm BSM} \simgt 500$\,GeV. 
Pair production of this particle in UHECR interactions with nucleons at rest would then occur only beyond a threshold of $\ecr \simgt 1$\,PeV. 
This threshold energy would be inherited (with some losses) by the BSM particle and its decay products, and so defines the approximate minimum energy of interest, before consideration of Earth propagation effects. 

Finally, BSM scenarios naturally anticipate the otherwise unlikely zenith angles of the AAEs. 
Specifically, the BSM particle's relatively long lifetime and high Lorentz factor will deplete event rates requiring observation of the daughter tau particles close to the horizon, while energy losses and decays too deep within the Earth for tau emergence will deplete tau-based event rates toward the nadir (event rates for the BSM particle will be greater near the horizon, but the particle itself is, by construction, difficult to detect and distinguish from backgrounds). 
While the exact resulting distribution of zenith angles will depend on the competition between these countervailing effects, and likely requires simulation to predict in detail, overall the BSM scenario provides a robust prediction of a maximal event rate, above any given energy threshold, at some intermediate zenith angle, $100\arcdeg \simlt z_{\rm max}(\etau) \simlt 150\arcdeg$. 
This is consistent with \anita\ observations for $z_{\rm max}(0.1\,{\rm EeV}) \approx 120\arcdeg$, the mean of the two AAE zenith angles. 

All of this may read as a compelling story; however, caution advises seeking independent confirmation before declaring discovery of a new BSM phenomenon. 
We therefore turn now to exploring the prospects for such confirmation via other facilities. 


\section{Insights from Other Observatories\label{sec:other}}

Having excluded SM explanations for the AAEs, we consider other facilities that may be able to provide independent confirmation or refutation of the inferred physical phenomenon of EeV-scale steeply-inclined upgoing cosmic rays.
We shall refer to the physical events, independent of detecting facility, as ``Sub-EeV Earth-emergent Cosmic Rays'' or SEECRs.

As a first step, we seek to anticipate event rates and fluxes for the SEECRs as they might be observed by other facilities.
The \anita\ energy calibration developed in \cite{sbr+16} enabled those authors to compare the rate of reflected cosmic rays (as a function of energy) to the well-known UHECR spectrum \cite{auger18_crspec}; along with a detailed simulation of the flight, this allowed them to estimate the effective energy-dependent exposure for \anitaone. 
They find an exposure of 1.2\,\kmsqsryr\ for reflected events of the mean observed energy of 2.9\,EeV. 
Observation of 14 reflected events for \anitaone\ \cite{sbr+16} and 17 for \anitathree\ \cite{gra+18} then suggests a total exposure of 2.7\,\kmsqsryr\ for the two flights together.

While the AAE energies are significantly lower than for the reflected events, they are nonetheless detectable because they achieve shower maximum near the surface, in high-density regions of the atmosphere, do not experience losses in reflecting off the ice surface, and traverse a substantially shorter path length before detection by \anita. 
We therefore adopt the above exposure to estimate the event rate implied by detection of two AAEs: $r_{\mathcal{S},{\rm A}} \approx 0.74$ SEECR \perkmsqsryr\ for $\ecr > 0.1$\,EeV.  
Until further details are forthcoming from the \anita\ team, it will be difficult to characterize the \anita\ exposure for SEECRs with greater precision.

This \anita-based event rate estimate serves as an approximate upper limit, which will be reduced if the angular coverage of \anita\ is better suited for detection of SEECRs than for detection of reflected UHECRs. 
The magnitude of this differential will require a full instrumental characterization and, most likely, model-specific BSM simulations to explore in detail. 
However, such a scenario is easy to imagine given the BSM picture sketched out above  (Sec.~\ref{sub:smexclude}), which leads to a concentration of SEECR events at intermediate zenith angles, where \anita\ has good sensitivity, and a marked deficit toward the nadir, where \anita\ has none \cite{anita09_instrument,hng+10}, and therefore misses a relatively large number of incident reflected UHECR radio pulses. 

While the Pierre Auger Observatory has observed cosmic rays over an area of 3000\,\kmsq\ for more than a decade, accumulating a total exposure of over 67,000\,\kmsqsryr\ \cite{auger18_crspec}, it has pursued UHE neutrino searches only over a narrow range of zenith angles ($58\fdg 5 < z < 95\arcdeg$) close to the horizon \cite{auger15_nu}. 
Within this range of zenith angles, the near-horizontal or upgoing nature of a tau neutrino-induced shower can, under appropriate circumstances, be distinguished from other cosmic ray events, for example, through the relative electromagnetic and muonic content of the air showers. 
No events satisfying the strict selection criteria were seen, leading to limits on the diffuse UHE neutrino flux cited previously \cite{auger15_nu}. 

We have considered whether upgoing air showers with larger (AAE-like) zenith angles could be observed by the Auger Observatory. 
These events could only be detected using a monocular reconstruction of the fluorescence detector (FD) data.
This is because SEECRs would not trigger enough surface detectors to satisfy the requirements to form a hybrid event for Auger.
If a new monocular reconstruction sequence were developed requiring an upgoing air shower with no signals in the surface array, it might be possible to identify AAEs, or set limits on their rate of occurrence, using the existing FD-only data. 
In this fashion, it may be possible to identify SEECRs as upgoing air showers in archival data of the Auger Observatory \cite{auger16_upgrade} or the Telescope Array \cite{ta08_instrument}.

The \icelong\ has been operating at its design sensitivity since 2010, and recently reported results of a nine-year (3142.5 day) search for PeV--EeV cosmogenic neutrinos \cite{ic18_diffnu}, including detection of two events with $\enu \simgt 1$\,PeV deposited energy. 
To estimate the total sensitivity of this search to SEECRs we note that, as with the \anita-detected showers, the key question is whether a high-energy $\tau$, or associated particle shower, is produced near enough to an appropriate boundary layer. 
In the case of \ice, the boundary is the instrumented volume of the detector, with a projected surface area (effectively independent of direction) of 1~\kmsq. 
In the case of \anita, the boundary is the Antarctic surface.
We thus estimate the associated \ice\ exposure for SEECRs events as follows: 
the one \kmsq\ projected area of \ice\ times the active exposure of the EHE search (8.6~years), times $2\pi$\,sr to account for the upgoing nature of the events.  
Multiplication of these factors gives an estimated \ice\ exposure of 54.0\,\kmsqsryr.

The \ice\ exposure is thus at most 20 times the \anita\ exposure, with perhaps three events seen (discussed below), leading to an order of magnitude discrepancy with the \anita\  event rate estimate. 
As stated, however, the \anita\ estimate is a model-dependent upper bound that is likely to be reduced somewhat under realistic model scenarios. 


\subsection{IceCube-140611\label{sub:icuptrk}}

Two $>$PeV neutrinos are identified in the nine-year \ice\ search reported in \cite{ic18_diffnu}. 
The first of these events, \icuptrk, is a throughgoing track with 2.6\,PeV deposited energy from ${\rm Dec.\ }+11\fdg 42$, also discussed in \cite{ic16_n_muon} (Table~\ref{tab:ice}). 
Interpreted as a muon track, it would have been produced by a significantly higher-energy ($\enu \approx 8.7$\,PeV, median expectation) neutrino, via either CC interaction of a muon neutrino (88\% chance), muonic decay of a tau neutrino (11\% chance), or decay of a $W^{-}$ produced by electron antineutrino interaction (1.4\% chance). 
The second event, from December 2016 (more precise timing, positional, and energy estimates are not yet available), is a partially-contained cascade with 2.7\,PeV deposited energy.
Cascade events can be produced by NC interactions of any flavor neutrino or by CC interactions of electron neutrinos; the inferred incident neutrino energy is $\enu \approx 5.9$\,PeV, making it a candidate Glashow resonance event. 


\begin{table*}
  \caption{Properties of IceCube Anomalous Track Events\label{tab:ice}} 
  \begin{ruledtabular}
    \begin{tabular}{lccc}
    \chead{Property} & \chead{IceCube-140611}
                     & \chead{IceCube-140109}
                     & \chead{IceCube-121205} \\ \hline
      EHE Northern Track ID & \#27 
                            & \#24 
                            & \#20 \\
      Date \& Time (UTC or MJD) & 2014-06-11 04:54:24 
                         & 56666.5
                         & 56266.6 \\
      Equatorial coordinates (J2000) & R.A. $110\fdg 34 \pm 0\fdg 22$,
                       & R.A. $293\fdg 29$,
                       & R.A. $169\fdg 61$, \\
      \mbox{~}         & Dec.\ $+11\fdg 42 \pm 0\fdg 08$ 
                       & Dec.\ $+32\fdg 82$
                       & Dec.\ $+28\fdg 04$ \\
      Zenith angle $z$ & $101\fdg 42$  
                       & $122\fdg 82$
                       & $118\fdg 04$ \\
      Earth chord length $\ell$ & 2535\,km 
                                & 6910\,km
                                & 5990\,km \\ \hline
      As muon: \emuobs\ (\eproxy) & 4.45\,PeV
                                  & 0.85\,PeV
                                  & 0.75\,PeV \\ 
      \phm{As muon:\,} $\enu$ (median) & 8.7\,PeV 
                                       & 1.65\,PeV
                                       & 1.45\,PeV \\
      \phm{As muon:\,} Mean interaction length for \enu                                            & 1960\,km 
                               & 3280\,km
                               & 3690\,km \\ 
      \phm{As muon:\,} \peveobs & $4.0\times 10^{-3}$ 
                                & $6.9\times 10^{-2}$ 
                                & $8.6\times 10^{-2}$ \\
      \phm{As muon:\,} \pzvzobs & $1.5\times 10^{-1}$
                                & $5.0\times 10^{-2}$
                                & $8.8\times 10^{-2}$ \\
      \phm{As muon:\,} \pjoint & $4.9\times 10^{-3}$ 
                               & $2.3\times 10^{-2}$
                               & $4.5\times 10^{-2}$ \\ \hline
      As tau: \etauobs\ (median) & 70\,PeV
                                 & 13\,PeV 
                                 & 12\,PeV \\
      \phm{As tau:\,} Mean interaction length for $\enu=1$\,EeV
                             & 340\,km
                             & 270\,km
                             & 285\,km \\
      \phm{As tau:\,} $p_{\rm SM}(\etau > \etauobs)$ for $\enu=1$\,EeV 
                             & $2.2\times 10^{-4}$ 
                             & $3.8\times 10^{-6}$
                             & $1.0\times 10^{-5}$ \\ 
      \phm{As tau:\,} $p_{\rm SM}(z > z_{\rm obs})$ for
                         $\enu=1$\,EeV, $\etau > \etauobs$
                             & $5.0 \times 10^{-3}$
                             & $4.5 \times 10^{-5}$
                             & $1.8 \times 10^{-4}$ \\
    \end{tabular}
  \end{ruledtabular}
\end{table*}


The relative likelihood for these events to be produced by cosmogenic (Berezinsky-Zatsepin or BZ) neutrinos \cite{bz69,aag+10,kao10,aal+18}, as compared to a $\varepsilon_\nu^{-2}$ extrapolation of the TeV--PeV diffuse astrophysical flux discovered by \ice\ \cite{ic13_astronu} and previously characterized over $30{\rm \,TeV} \simlt \enu \simlt 2{\rm \,PeV}$, leads \ice\ to conclude that both neutrinos are likely astrophysical in origin \cite{ic18_diffnu}.

Attribution of \icuptrk\ to an $\varepsilon_\nu^{-2}$ astrophysical flux at $\enu>5$\,PeV energies may be premature, however, for reasons explored by \cite{kl18}.
First, evaluation of the likelihood under models for cosmogenic neutrinos will underestimate the odds of its originating in a distinct spectral component peaking at energies more appropriate to the observed energy of the event. 
Second, the relatively soft neutrino spectrum inferred from cascades, the only calorimetric events for \ice, extrapolates to very low event rates at $\enu > 8$\,PeV; \cite{kl18} find $r_{\rm t} \approx 0.026$\,\peryr\ for $\enu > 5$\,PeV, with a 2:1 ratio of downgoing ($80\arcdeg < z < 90\arcdeg$) to upgoing events.
Third, evading these constraints with a harder spectrum, as suggested by analysis of high-energy muon tracks \cite{ic16_n_muon}, may (depending on source models) run into the separate challenge of conflicting with the paucity of observed Glashow resonance events at 6.3\,PeV \cite{glashow60}.  

As their alternative hypothesis, \cite{kl18} suggest \icuptrk\ may be a misidentified upgoing tau track with energy $\etau \approx 0.07$\,EeV -- in other words, a SEECR (Table~\ref{tab:ice}).
In support of this hypothesis they note the following: First, while tau leptons are capable of producing distinct detector signatures including ``lollipop'' and ``double bang'' events \cite{lp95,bbh+03,dyrc07}, for a km-scale detector like \ice\ these signatures are most evident at energies $\enu < 20$\,PeV where the mean distance to tau decay is $\gamma c \tau_\tau < 1$\,km.
Above this energy a majority of tau detections manifest as simple tracks, with little to distinguish them from muon tracks of $m_\tau/m_\mu \approx 17$ times lower energy. 
Second, assertion of a new spectral component at sub-EeV energies, as an explanation for \icuptrk, nicely resolves the tension between the event's inferred energy under a muon interpretation and the soft spectrum of the TeV--PeV diffuse astrophysical neutrinos and absence of Glashow events, as already noted.

Two unresolved questions highlighted by \cite{kl18} are, first, why \icuptrk\ is observed at a relatively large zenith angle which ought to be disfavored due to Earth opacity effects (see their Fig.~3); and second, what underlying neutrino source population(s) would generate a tau neutrino with 70\,PeV energy (they explore the consequences of three toy models).

The second question must be answered for any model that seeks to explain the SEECR phenomenon. 
We note that the nature of the SEECRs as due to $\etau \simlt 0.1$\,EeV tau leptons, generated by interactions of $\enu \simlt 1$\,EeV tau neutrinos, already provides a reasonable match to the observed properties of the AAEs. 
In this sense, associating the two classes of event is natural; almost any explanation for one will also suffice for the other. 
We discuss theoretical precedents for these events in Sec.~\ref{sec:theory}. 

The first question will be familiar from our exploration of the \anita\ anomalous events, above. 
As with the AAEs, then, we explore SM expectations for UHE tau neutrinos incident along the trajectory of \icuptrk\ using \nutausim. 
We inject 100~million tau neutrinos at each half-decade of energy between 0.1\,EeV and 1000\,EeV (nine energies in all), and tabulate the energies of all emerging tau particles with $\etau > 0.1$\,PeV. 
Since the chord length for \icuptrk\ is significantly shorter than those of the AAEs, the break in the energy distribution above $\etau \approx \ecritt$ is not as sharp. 
The maximum observed tau energy is almost 10\,EeV, resulting from propagation of an $\enu = 100$\,EeV tau neutrino, and the greatest fraction of $\etau > 3.2$\,EeV emergent tau are generated by 10\,EeV incident tau neutrinos. 
Overall, we find a fraction of $2.2\times 10^{-4}$ incident $\enu = 1$\,EeV tau neutrinos successfully yield an emergent tau with $\etau > 0.07$\,EeV, the most likely tau energy for \icuptrk\ \cite{kl18}. 
With regard to the zenith angle distribution over $82\fdg 5 < z < 180\arcdeg$ (the angular range probed for such events by \cite{ic18_diffnu}), we find the $p$-value for a 1\,EeV tau neutrino yielding a $\etau > 0.07$\,EeV tau to be observed from $z>101\fdg 42$ is $\pzvzobs = 5.0\times 10^{-3}$ (Table~\ref{tab:ice}).

Completely independent of \anita\ observations, then, a tau track interpretation for \icuptrk\ is in 2.8$\sigma$ tension with the SM, due to its unexpectedly large zenith angle.

This does not mean that the tau track interpretation of \icuptrk\ is correct. That case has been made by \cite{kl18}, although without any final characterization of confidence level. 
Our own attempt to answer this question is presented, as part of our analysis of the \ice\ Extremely High Energy Northern track events, in the next section. 


\subsection{IceCube Northern Track Events\label{sub:pnts}}

\ice\ has observed a persistent tension between the harder astrophysical neutrino spectrum (neutrino index $\gamma=2.19\pm 0.10$ \cite{ic17_icrc_astro}) measured from primarily northern track events (hereafter PNTs; \cite{ic15_n_muon,ic16_n_muon,ic17_icrc_astro}), and the softer spectrum ($\gamma=2.92^{+0.33}_{-0.29}$ \cite{ic17_icrc_astro}) measured from the all-sky HESE (High Energy Starting Event; including cascades and tracks) sample \cite{ic13_astronu,ic14_hese,ic17_icrc_astro}.  
The tension between HESE and PNT spectra, nominally $>$2.3$\sigma$, is reduced to the $\approx$95\% confidence level when considering correlated uncertainties between power-law index and normalization \cite{ic17_icrc_astro}.

It is possible that independent support for disagreement between the two analyses may be found in the absence of Glashow events: 
Using the PNT spectrum from \cite{ic16_n_muon}, \cite{kl18} find that the comparable (reduced by 50\% in normalization) unbroken electron antineutrino spectrum is excluded at $\approx$99\% confidence by the absence of Glashow events. 
However, this conclusion depends on neutrino source models, since $p\gamma$ production (as likely for the only known neutrino-emitting blazar, TXS~0506+056 \cite{kmp+18}) yields a minimal flux of electron antineutrinos, and hence very few Glashow events.  

Given the strength of the arguments in favor of interpreting \icuptrk\ as a misidentified tau track event, it is worth considering the impact such a misidentification would have upon analysis of the larger PNT sample. 
\icuptrk\ is one of just 36 events in the ``Extremely High Energy'' (EHE; $\emu > 200$\,TeV) portion of the PNT sample \cite{ic16_n_muon,ic17_icrc_astro}, and its estimated energy is more than 4.5 times greater than that of any other PNT event. 
Misidentification of this single event, then, and its use in characterizing the incident neutrino spectrum at $\enu \sim 10$\,PeV energies (as in the muon track interpretation) rather than at $\enu \sim 0.1$\,EeV energies (as in the tau track interpretation), would be expected either to lead to a significant hardening of the inferred spectral index for the PNT analysis, or to render a single power-law fit to the dataset unacceptable. 

To the contrary, PNT spectral fits have remained consistent before and after inclusion of \icuptrk\ in the sample, and investigations have found no evidence for deviation from a single power-law fit across the full set of PNT neutrino energies \cite{ic16_n_muon,ic17_icrc_astro}. 
This would be puzzling, unless \icuptrk\ were not the only misidentified tau track among the PNTs.
Since the astrophysical signal is most strongly distinguished from atmospheric background using $\enu > 120$\,TeV neutrinos \cite{ic17_icrc_astro}, hardening of the spectrum can occur via contamination by further misidentified tau tracks above the median neutrino energy beyond this threshold (for a $\gamma=2.2$ spectrum), $\enu > 210$\,TeV. 
At these energies, the inferred tau particle would have 17$\times$ greater energy, $\etau \simgt 3.6$\,PeV. 
This is reasonably proximate to the $\etau \simgt 20$\,PeV energy required for the tau to travel more than 1\,km and thus (typically) appear as a simple track in \ice. 
Since these hypothetical high-energy tau events are present only in the PNT sample, and have energies $>$0.01\,EeV, they would represent a further, previously unidentified population of SEECRs.

\ice\ maintains a catalog of $\emu > 200$\,TeV EHE northern track events (hereafter EHENTs; \cite{ic16_n_muon,ic17_icrc_astro}), now numbering 36 including \icuptrk. 
We can thus attempt to discern the presence of possible ``hidden SEECRs'' among these events. 

We make a global analysis of the EHENT energies and zenith angles \cite{ic16_n_muon,ic17_icrc_astro}. 
Inspired by the properties of \icuptrk\ and the AAEs, we focus on two distinguishing features of the SEECRs as compared to the lower-energy muons that will provide the dominant component of the EHENT sample. 
First, we hypothesize that even though they have been misidentified as lower-energy muon tracks, the SEECRs nonetheless provide the higher-energy events of this sample, and are thus responsible for hardening the PNT spectrum compared to HESE analyses. 
Second, since the AAEs are observed from ``impossible'' zenith angles, significantly larger than expected in the SM, we expect analogous EHENT SEECRs to be incident from unlikely (large) zenith angles (again, even though the event energy has been underestimated).

Both analyses will benefit from a clear understanding of the likely incident neutrino energy for each EHENT event. 
The highly stochastic nature of high-energy muon (or tau) propagation in ice, with a substantial fraction of all light production associated with individual nuclear scattering bremsstrahlung events, means this inference is unavoidably uncertain. 
Simulations of the \icuptrk\ event \cite{ic16_n_muon} indicate that the median neutrino energy in a muon track interpretation is 1.94 times the catalog muon energy value \eproxy. 
We adopt this median correction factor to estimate the incident neutrino energies of EHENT events under the null hypothesis. 

The EHENT energy analysis calculates $p$-values for each event on the basis of its energy as \peveobs. 
As an incident neutrino spectrum we use the recent \ice\ four-year pure-cascade analysis \cite{ic17_icrc_astro}, which reports best-fit values for the conventional atmospheric and astrophysical contributions to the observed neutrino spectrum (neutrino index $\gamma=2.48\pm 0.08$ for the astrophysical component).  
This spectral analysis will be free of contamination from misidentified tracks and offers the highest-quality calorimetry of any \ice\ sample. 
For each EHENT, we integrate the \ice\ sky coverage over the relevant zenith angles ($z\ge 90\arcdeg$ for events \#1 and \#2; $z\ge 85\arcdeg$ for the remaining events) for neutrino energies $388\,{\rm TeV} \le \enu \le 100\,{\rm PeV}$, weighting by the cascade analysis atmospheric + astrophysical spectrum and the transmission probability for neutrinos of each energy (as a pure exponential in interaction depth, using the reference Earth model and median neutrino cross sections from \cite{acp+18}). 
(We note that we approximate the atmospheric neutrino flux, a minority component at these energies, as isotropic although it is not \cite{cr03}.)
We convert this PDF to a CDF and determine $p$-values \peveobs.

The EHENT zenith angle analysis calculates \pzvzobs\ for each event on the basis of its zenith angle conditioned on its energy. 
With the energy of each event assumed, this is an independent statistic from the \peveobs\ values already calculated. 
Treating neutrino transmission probability as an exponential with interaction depth, and using the reference Earth model and median neutrino cross sections from \cite{acp+18}, we integrate the transmission probability over the relevant zenith angles for each event and normalize to get a PDF for zenith angle, given the neutrino energy. 
The $p$-values \pzvzobs\ are then calculated by integration of this PDF.

\begin{figure}
  \begin{centering}
    \includegraphics[width=0.45\textwidth]{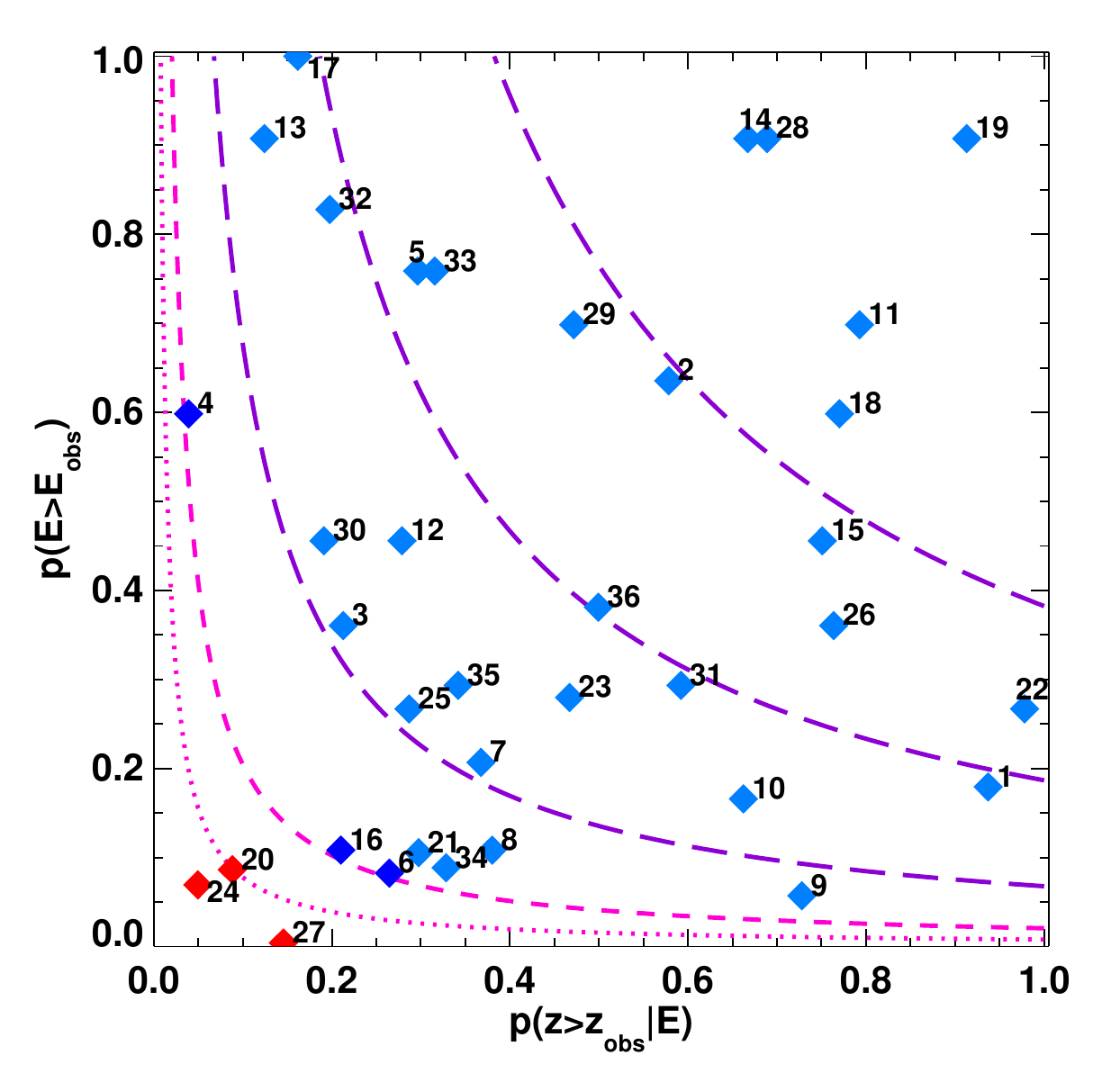}
  \end{centering}
\caption{Energy-based \peveobs\ and zenith angle-based \pzvzobs\ $p$-values for the 36 extremely high-energy northern track (EHENT) events from \ice\ \cite{ic16_n_muon,ic17_icrc_astro}. 
  Event ID numbers are indicated, and contours of fixed joint probabilities of 4.6\% (2$\sigma$; dotted magenta), 10\% (dashed magenta), 25\%, 50\%, and 75\% (long dashed purple) are shown. 
  Event \#27 is \icuptrk, with $\pjoint = 0.49\%$; in addition to \icuptrk, events \#24 ($\pjoint = 2.3\%$) and \#20 ($\pjoint = 4.5\%$) also lie beyond the 2$\sigma$ contour (red symbols). 
  Events \#4 ($\pjoint = 10.5\%$), \#6 ($\pjoint = 10.9\%$), and \#16 ($\pjoint = 11.2\%$), plotted with dark blue symbols, lie just within the 90\%-confidence contour. \label{fig:ehe_p2d}}
\end{figure}

Results of these two analyses are shown in Fig.~\ref{fig:ehe_p2d}, which plots \pzvzobs\ versus \peveobs\ for the 36 EHENT events \cite{ic16_n_muon,ic17_icrc_astro}. 
Event ID numbers are indicated, and contours of fixed joint probability (4.6\% or 2$\sigma$; 10\% or 90\%-confidence; and 25\%, 50\% and 75\%), derived by Fisher's method from the two $p$-values for any point, are shown for reference purposes.


As regards the population, the joint analysis provides support at $>$90\% confidence against the null hypothesis that the EHENTs are produced by muon tracks generated by astrophysical (and minority atmospheric) neutrinos with spectra as measured in the cascade analysis; this is consistent with the observation that the PNT and cascade spectral fits are incompatible at 96\% confidence \cite{ic17_icrc_astro}. 
Quantitatively speaking, the full set of 72 $p$-values yields $\pjoint=8.7\%$ by Fisher's method (91.3\% confidence), while the component energy and zenith angle analyses (36 $p$-values each) give $p_\varepsilon=6.9\%$ and $p_z=32\%$, respectively. 
For both the energy and zenith angle analyses, the sense of the deviation is what would be expected for a mild contamination of the EHENT sample by misidentified tau tracks, as yet insufficient to demonstrate the presence of such contamination at high confidence.

Nonetheless, we can use these analyses to quantify the significance of \icuptrk, and to identify additional EHENT events that exhibit the most improbable energies and zenith angles under the null hypothesis. 

For \icuptrk\ (\#27), we find $\pjoint=0.49\%$, which corresponds to 2.8$\sigma$ confidence for a single trial. 
In isolation, a muon track origin for this event is thus strongly disfavored, consistent with the intuition expressed (albeit unquantified) by \cite{kl18}. 
Indeed, \icuptrk\ is a singular event, the only EHENT event to satisfy the selection cuts for the \ice\ PeV--EeV diffuse neutrino flux analysis \cite{ic18_diffnu}, and the second-highest $E_{\rm dep}$ event yet observed by \ice; moreover, it is incident from an improbably large zenith angle for its energy, $\pzvzobs=11.5\%$. 
That said, it must be acknowledged that the strength of exclusion for the muon track hypothesis for this event, after trials correction for the 36 EHENT events, is just 16\% or 1.4$\sigma$. 

Beyond \icuptrk, which remains the most interesting (improbable under a muon track interpretation) single event in our analysis, we identify two further candidate SEECRs in the lower-left (smallest \pjoint) region of Fig.~\ref{fig:ehe_p2d}, beyond the 2$\sigma$ contour (dotted magenta line). 
These events are: \icuptwo\ (\#24) with $\eproxy=850$\,TeV, ${\rm Dec.}  +32\fdg 82$, and $\pjoint=2.3\%$; and \icupthree\ (\#20) with $\eproxy=750$\,TeV, ${\rm Dec.} +28\fdg 04$, and $\pjoint=4.5\%$. 
We list these events, with their observed and calculated properties under both muon and tau track interpretations, alongside \icuptrk\ in Table~\ref{tab:ice}. 

We carried out \nutausim\ simulations for the trajectories of both \icuptwo\ and \icupthree\ to determine the SM probability for a successful emergent ($\etau>\etauobs$) tau for each event.
We find $p_{\rm SM}(\etau>\etauobs)$ values of $3.8\times 10^{-6}$ (\icuptwo) and $1.0\times 10^{-5}$ (\icupthree), respectively. 
We then used simulations across the full range of zenith angles to evaluate the SM probability \pzvzobs\ under the tau track interpretation, finding $p$ values of $4.5\times 10^{-5}$ (\icuptwo) and $1.8\times 10^{-4}$ (\icupthree). 
Hence, as with \icuptrk, confirmation of these events as tau tracks would lead directly to a confrontation with SM expectations not too dissimilar from that presented for the AAEs in Sec.~\ref{sec:sm}. 
We therefore find that these two events are \ice\ anomalous track events -- not just candidate misidentified tau tracks, but candidate SEECRs. 

Notably, exclusion of the three events in Table~\ref{tab:ice} from the EHENT sample is required to yield a distribution in energy and zenith angle for the remaining EHENTs ($n=33$) that is consistent ($p=50\%$ for the combined analysis) with the cascade-only spectrum and normalization, and thus resolves the longstanding tension between the PNT analysis and other samples. 
This validates our earlier suspicion, demonstrating that contamination of the EHENT sample by just three misidentified tau tracks can cause $>$90\% tension with the cascade results. 
(We are not able to explore the consequences of excluding these events from the full PNT analysis, as data for the non-EHENT events in the PNT sample have not been released.) 

We note that $p$-values from the \peveobs\ component of our analysis will be biased low if there exist distinct astrophysical neutrino flux(es) which dominate over extrapolations of the known TeV--PeV astrophysical flux over the 10\,PeV to 100\,PeV energy range. 
No such component has yet been demonstrated; hypothesized source populations that might yield neutrino spectra peaking in this energy range include active galactic nuclei \cite{mid14,fm18}, magnetars \cite{mmz09,fkm+14}, gamma-ray burst early afterglows \cite{wb00,murase07,ry15} including UV/\xray\ flares \cite{mn06}, or dark matter decay processes \cite{mb12,sb18}. 

We emphasize that the statistical strength of our analysis is not sufficient to demonstrate a SEECR nature for any individual EHENT event, nor for the three \ice\ anomalous track events together. 
Rather, if detailed analyses of \ice\ data can provide independent evidence for a tau track interpretation for any of these events, then -- given the improbability of observing them as tau tracks under the SM (Table~\ref{tab:ice}) -- this would confirm their nature as SEECRs. 
Given the stochastic nature of the light output for extremely relativistic track events, it may be possible to construct a statistic that is sensitive to particle mass, independent of Lorentz factor, which could be used for this purpose. 

Confirming the existence of SEECRs in \ice\ data would validate the SEECRs as a physical phenomenon, support a SEECR interpretation of the AAEs, and demonstrate the existence of a responsible BSM particle. 
It would also clarify the nature of the SEECR particle as a UHE tau.
This has been suspected for the AAEs, but would be difficult to demonstrate with \anita\ data in the absence of other observations. 


\subsection{Support from IceCube\label{sub:support}}

Supporting observations from \ice\ for the reality of the SEECR phenomenon consist of the following. 

First, the spectral fit to the eight-year PNT sample \cite{ic17_icrc_astro}, which is in tension at $>$95\% confidence with the spectral fits derived from the four-year pure cascade and six-year HESE analyses \cite{ic17_icrc_astro}, and (depending on source physics) may also conflict with the absence of published Glashow resonance events \cite{kl18}. 
While a new and softer Galactic component to the astrophysical neutrinos has been suggested as a possible cause of this discrepancy, this is disfavored by Galactic Center versus anti-Center comparisons \cite{ic16_n_muon}. 
By contrast, we have shown that a relatively mild contamination of the sample by misidentified tau tracks is sufficient to explain the discrepancy. 

Second, the singular event \icuptrk, first called out as a possible misidentified tau track by \cite{kl18}: we find that a muon track interpretation for this event is disfavored at 2.8$\sigma$, single trial (Table~\ref{tab:ice}). 

Third, we identify two further candidate SEECRs within the \ice\ EHENT sample, \icuptwo\ and \icupthree, for which a muon track interpretation is disfavored at $>$2$\sigma$, single trial (Fig.~\ref{fig:ehe_p2d}; Table~\ref{tab:ice}). 
Exclusion of these two events, and \icuptrk, from the EHENT sample results in a 33-event sample that is fully consistent with the energy distribution predicted from the four-year pure cascade analysis, and with the zenith angle distribution expected for muon neutrinos of the expected energies. 

Two recent \ice\ analyses \cite{ic17_icrc_astro,ic18_inelastic} have attempted to reconcile the PNT and HESE/cascade spectral results, either within a single spectral model or by identifying evidence for a spectral hardening or second component to the astrophysical neutrino spectrum, without success. 
In both cases, analysis of non-PNT track events is restricted to ``starting track'' events where the neutrino interaction yielding the relativistic muon is contained within the \ice\ detector array. 
Such starting track events are expected to be free of misidentified tau tracks, as the 17$\times$ higher energy interaction needed to produce a comparable tau track would be accompanied by a comparably higher-energy hadronic cascade at the interaction vertex. 
These analyses thus provide further support for the hypothesis that contamination by tau misidentification via SEECRs explains the discrepant PNT spectrum, since this is the only model that yields a substantially harder spectrum among the northern throughgoing track events that are exclusive to the PNT sample. 

We note that the Dec.~2016 partially contained shower event with $E_{\rm dep}=2.7$\,PeV ($\enu\approx 5.9$\,PeV) \cite{ic18_diffnu} may be a SEECR. 
Adopting the softer HESE/Cascade spectrum reduces the expectation for \ice\ to observe such a high-energy cascade event from the TeV--PeV astrophysical neutrinos. 
Tau neutrinos, however, can produce a cascade upon either production or decay of the resulting tau, in the ``double bang'' event topology \cite{lp95}. 
Depending on the incident direction of the neutrino, the observed event could either reflect the hadronic shower associated with the tau creation interaction, or the shower produced by hadronic or electromagnetic decay of the tau itself. 
Given the deposited energy, the incident tau neutrino energy would then be $\enu\approx 10.8$\,PeV (tau creation) or $\enu\approx 5.4$\,PeV (tau decay), depending on the nature of the observed cascade.
Then, if its zenith angle is sufficiently steep, the combination of properties would make this event a SEECR. 

Looking ahead, the most straightforward way to strengthen the \ice\ support for the SEECRs would be to carry out a joint (multi-sample) spectral analysis that explicitly allows for tau track misidentification, and a population of $5\,{\rm PeV} \simlt \etau\ \simlt 0.5\,{\rm EeV}$ tau particles, within the northern throughgoing track sample only. 

A somewhat more challenging task, proposed by \cite{kl18} and reiterated by us above, would be to develop an analysis yielding mass constraints for relativistic charged leptons from the statistical patterns of light deposition observed by \ice.  
While difficult, if it succeeds, such an analysis would provide valuable event by event discrimination of muon and tau tracks, and enable individual identification of high-confidence SEECRs. 

Finally, the presence of three SEECRs in the eight-year PNT sample \cite{ic17_icrc_astro} would give an independent rate estimate of $r_{\mathcal{S},{\rm I}} \sim 0.06$\,SEECR \perkmsqsryr\ over \ice's upgoing hemisphere. 
As mentioned previously (Sec.~\ref{sec:other}), this is roughly an order of magnitude below the event rate upper limit estimated from \anita\ observations of the AAEs, before accounting for any underlying event spectrum (the \ice\ events are very likely lower energy than the AAEs). 

This discrepancy between event rates is model-dependent, and must remain unresolved until comprehensive simulations can explore the interplay between astrophysics, particle physics, and facility / instrumental effects that lead to successful production, propagation, and detection of SEECRs of various energies, incident from various angles. 

We note one possible observational bias that we have not seen explicitly addressed by the \anita\ team. 
While they discuss an extensive suite of \texttt{ZHAireS} \cite{acr+12} simulations of the reflected UHECR showers they observe \cite{sbr+16}, they do not mention simulating upgoing events (AAEs) with the same fidelity -- rather, reported AAE properties are inferred by analogy to similar reflected events \cite{gra+18}. 
Yet development of the UHECR air shower and  properties of the resulting radio pulse may be significantly affected by propagating down a density gradient (as for upgoing events only), rather than up. 
In particular, if the resulting beam of bright radio emission is significantly broadened by comparison to the downgoing events, this would lead to overestimated event rates from \anita. 
It might also result in underestimation of event energies; since we have treated event energies conservatively, we do not expect that our analysis (Sec.~\ref{sec:sm}) would be significantly affected. 


\section{Theoretical Precedents\label{sec:theory}}

We will now review some theoretical precedents for the SEECR phenomenon from the literature. 

Model-independent constraints on the properties of the hypothetical BSM particle required to explain the existence and properties of the SEECRs are roughly as follows: 
The particle should couple to the tau or tau neutrino. 
Its cross section for nuclear interactions at $\sim$EeV energies should be one or two orders of magnitude less than the total neutrino cross section of 15\,nb at 1\,EeV, so that its mean free path through the Earth is $>$1000\,km, while allowing for a reasonable branching ratio in UHECR neutrino + nucleon interactions \cite{abc04,abp+08}. 
Its lifetime should be of order $\tau_{\rm BSM} \sim 10\,(m_{\rm BSM}/500\,{\rm GeV})$\,ns, so that at EeV energy it propagates roughly an Earth radius. 

Remarkably, particles satisfying these criteria are anticipated within existing gauge-mediated supersymmetry breaking (GMSB) supersymmetric (SUSY) models of the fundamental interactions.
Under these models, and across a range of GMSB model parameters, the next-to-lowest mass supersymmetric particle (NLSP) is a relatively long-lived stau (\stau) which can be produced with reasonable branching ratios (${\rm BR} \simlt 10^{-4}$) via UHECR neutrino + nucleon interactions \cite{abc04,abp+08}. 
Given anticipated \stau\ + nucleon cross sections of 100~pb or less \cite{abc04,abp+08} and minimal ionization, bremsstrahlung, pair production, and photo-nuclear energy losses \cite{abc07,abp+08}, the stau then propagates across much of the Earth and, for appropriate energies and lifetimes, decays to a tau lepton plus unseen stable lowest-mass supersymmetric particle (LSP) prior to Earth emergence \cite{abc04,akr06,abc07,aim+07,abp+08,ccmp09,ac12,ac13,cab18}. 

These models thus offer a precedent for SEECR observations. 
Indeed, the cited discussions include extensive treatments of potential observational consequences including ``impossible'' upgoing UHECR air showers \cite{ac13,cab18}, PeV--EeV charged particles incident on neutrino facilities from improbably large zenith angles \cite{abp+08,ac12}, and parallel ``pair tracks'' in neutrino facilities, produced by the two stau from a single pair production interaction \cite{abc04,akr06,aim+07,abp+08}. 
The first two of these anticipate the observed properties of the AAEs (upgoing UHECR air showers) and the \ice\ candidate SEECRs (upgoing PeV--EeV charged particles) very well; the last ``pair track'' observable is not expected in models where most \stau\ decay before reaching the detector -- stochastic losses mean the path lengths for the two resulting $\tau$ particles differ enough that there is little probability of observing both tracks. 

Within these scenarios, the stau mass should be within the range of $0.5\,{\rm TeV} \simlt m_{\mstau} \simlt 1.0\,{\rm TeV}$ to evade detection in completed LHC searches \cite{atlas18_dilepton,cms18_tau} while successfully resolving the SM hierarchy problem. 
Its associated LSP is then a stable $m_{\rm LSP} < m_{\mstau}$ particle with minimal interactions that can offer an attractive candidate for the dark matter via ``SuperWIMP'' scenarios (e.g., \cite{feng10}).


\section{Conclusions\label{sec:conclude}}

We have demonstrated (Sec.~\ref{sec:sm}) that SM models cannot explain the \anita\ anomalous events (AAEs) if they are correctly interpreted as $\ecr \approx 0.6$\,EeV upgoing UHECR showers with exit angles of $27\arcdeg$ to $35\arcdeg$, or as we have termed them, Sub-EeV Earth-emergent Cosmic Rays (SEECRs).

Having excluded SM explanations at $>$5$\sigma$ confidence, we have argued that the SEECR phenomenon, if confirmed, can support a discovery-level claim of ``beyond the Standard Model'' (BSM) physics. 

We have argued (Sec.~\ref{sec:other}) that existing supporting observations from \ice\ -- namely, the otherwise puzzling tension between the astrophysical neutrino spectrum inferred from primarily northern track events versus other samples, the singular event \icuptrk, and our own identification of two further SEECR candidates in the \ice\ $\eproxy > 200$\,TeV northern track sample -- provide reason to think this will happen soon. 
Confirmation of the reality of the SEECR phenomenon would in turn demonstrate the existence of a responsible BSM particle.

We have reviewed (Sec.~\ref{sec:theory}) the theoretical precedents \cite{abc04,akr06,abc07,aim+07,abp+08,ac12,ac13} that have anticipated the SEECR phenomenon, predicting observations of upgoing UHECR air showers \cite{ac13,cab18} and anomalous low-ionization track events in high-energy neutrino observatories \cite{abp+08,ac12} under suitable theoretical scenarios. 
These investigations were motivated by the observation that, across a range of model parameters for some supersymmetric (SUSY) theories of the fundamental interactions, the next-to-lowest mass SUSY partner particle is a relatively long-lived stau slepton (\stau) that can be produced in UHECR neutrino + nucleon interactions, and propagate much of the way through the Earth with minimal energy losses, before decaying to a PeV--EeV tau. 
The existence of such a BSM particle is the key missing element needed to resolve, in self-consistent fashion, the various improbable aspects of the AAEs and anomalous \ice\ track events, which otherwise make existence of these events (as a set) extremely unlikely. 

Within these SUSY-motivated scenarios, the SEECR particle's associated least-mass SUSY particle can provide an attractive candidate for the dark matter \cite{feng10}.


Looking forward, the most urgent priority will be to build the sample of robust and well-characterized SEECRs by a wide variety of means. 
Further balloon-borne searches, having proven their utility, will be eagerly anticipated, including the recently completed \anitafour\ mission, now in a data analysis phase. 
Data from \anitafour\ might well include new AAE SEECRs, and a further flight is planned \cite{pnd+18}. 
Future flights of the \eusospbtwo\ fluorescence telescope balloon may also be capable of identifying and characterizing SEECRs \cite{euso17_spb2}. 

Simultaneously, the kilometer-scale high-energy neutrino facilities will continue with data collection (\ice) and deployment (\kmnet), and further SEECR detections from \ice\ and initial data from \kmnet\ can be anticipated. 
Since these facilities retain high sensitivity and low background down to PeV scales, with sufficient exposure they should be able to measure the spectrum of the SEECRs. 
Associated theoretical and simulations work should seek to define additional observational signatures of particle mass for high-energy track events in these facilities, and \kmnet\ may wish to consider supplementing their detector array with temperature or infrared sensors to provide an independent means of mass discrimination for track events. 

The \jemeuso\ space experiment \cite{kem+11} could provide very large exposures once on-orbit, $>$600,000\,\kmsqsryr\ per year, but its planned threshold energy of 10\,EeV \cite{euso13_jemeuso} would prevent detection of SEECRs.

SEECRs may exist in archival fluorescence telescope data of UHECR experiments including the Pierre Auger Observatory \cite{auger16_upgrade} and the Telescope Array \cite{ta08_instrument}.
The challenge here will be to efficiently identify upgoing events, while rejecting downgoing air showers that may mimic the time structure of an upgoing event.


Finally, validity of SUSY-based models for the SEECRs will be tested in future runs of the Large Hadron Collider \cite{eb08} via its ATLAS \cite{atlas08} and CMS \cite{cms08} detector collaborations. 
If the SEECRs are correctly attributed to decay of SUSY particles with masses $0.5\,{\rm TeV} \simlt m_{\mstau} \simlt 1.0\,{\rm TeV}$, then a rich array of supersymmetric phenomena lie within reach of this facility, awaiting discovery. 


\begin{acknowledgments}

  The authors acknowledge useful conversations with members of the ANITA team, including A.~L. Connolly and J.~J. Beatty. 
  The authors gratefully acknowledge support from Penn State's Office of the Senior Vice President for Research, the Eberly College of Science, and the Institute for Gravitation and the Cosmos. 
  This work was supported in part by the National Science Foundation under grant \mbox{PHY-1708146}. 
  Portions of this work were carried out at the Aspen Center for Physics, which is supported by National Science Foundation under grant \mbox{PHY-1607611}.
  K.M. is supported by the Alfred P. Sloan Foundation and by the National Science Foundation under grant \mbox{PHY-1620777}.
  
\end{acknowledgments}


%



\end{document}